\documentclass[12pt,draftcls,onecolumn]{IEEEtran}

\usepackage{epsf}
\usepackage{graphics}
\usepackage{times}
\usepackage{graphicx}
\usepackage{amsmath}
\usepackage{amssymb}
\usepackage{epsfig}
\usepackage{subfigure}
\usepackage{array}
\usepackage{float}

\begin{document}

\title{A New Low-Complexity Selected Mapping Scheme Using Cyclic Shifted IFFT for PAPR Reduction in OFDM Systems}

\author{Kee-Hoon Kim, Hyun-Bae Jeon, Jong-Seon No, and Dong-Joon
Shin

\thanks{K.-H. Kim, H.-B. Jeon, and J.-S. No are with the
Department of Electrical Engineering and Computer Science, INMC, Seoul
National University, Seoul, 151-744, Korea (email: kkh@ccl.snu.ac.kr, lucidream@ccl.snu.ac.kr, jsno@snu.ac.kr).}
\thanks{D.-J. Shin is with the Department of
Electronic Engineering,
Hanyang University, Seoul, 133-791, Korea (email: djshin@hanyang.ac.kr).}
\thanks{A part of this paper was presented in IEICE International Symposium on Information Theory and its Applications \cite{Kim}.}
}

\maketitle

\begin{abstract}
In this paper, a new peak-to-average power ratio (PAPR) reduction scheme for orthogonal
frequency division multiplexing (OFDM) is
proposed based on the selected mapping (SLM)
scheme. The proposed SLM scheme generates alternative OFDM signal sequences by cyclically shifting the connections in each subblock at an intermediate stage of inverse
fast Fourier transform (IFFT). Compared with the conventional SLM scheme, the proposed SLM scheme achieves similar PAPR
reduction performance with much lower computational
complexity and no bit error rate (BER) degradation. The performance of the proposed SLM scheme is verified through numerical analysis. Also, it is shown that the proposed SLM scheme has the lowest computational complexity among the existing low-complexity SLM schemes exploiting the signals at an intermediate stage of IFFT.
\end{abstract}


\section{Introduction}

\IEEEPARstart{O}{rthogonal} frequency division multiplexing (OFDM) is a multicarrier modulation
method utilizing the orthogonality of subcarriers. OFDM has been
adopted as a standard modulation method in several wireless communication systems such as digital audio broadcasting (DAB),
digital video broadcasting (DVB), IEEE 802.11 wireless local area network (WLAN),
and IEEE 802.16 wireless metropolitan area network (WMAN).
Similar to other multicarrier schemes, OFDM has a high peak-to-average
power ratio (PAPR) problem, which makes its straightforward implementation quite costly.
Thus, it is highly desirable to reduce the PAPR of OFDM signals.

Over the last decades, various techniques to reduce the PAPR of OFDM signals have been proposed such as clipping~\cite{Oneal},\cite{YWang}, coding~\cite{Tsai}, active constellation extension (ACE)~\cite{Krongold}, tone reservation (TR)~\cite{Tellado}, partial transmit sequence (PTS)~\cite{muller},\cite{Yang}, and selected mapping (SLM)~\cite{Bauml},\cite{Goff}.
Among them, SLM and PTS are widely used because they show good PAPR reduction performance without bit error rate (BER) degradation.
However, they require many inverse fast Fourier transforms (IFFTs), which cause high computational complexity, and also need to transmit the side information (SI) delivering which phase rotation vector was used.

It is well known that SLM scheme is more advantageous than PTS scheme if the amount of SI is limited. However, the computational complexity of SLM scheme is larger than that of PTS scheme.
Therefore, many modified SLM schemes with low complexity have been proposed~\cite{Wang0}-\cite{Jeon}, but they have several shortcomings such as degradation of PAPR reduction performance or BER degradation compared to the conventional SLM scheme using the same number of alternative OFDM signal sequences. For example, Wang and Ouyang \cite{Wang0} proposed a low-complexity PAPR reduction algorithm but the elements of most phase rotation vectors have different magnitudes, which causes BER degradation. The low-complexity PAPR reduction algorithm in \cite{Wang} causes degradation of PAPR reduction performance because the used phase rotation vectors have periodicity and thus they are highly correlated. The scheme in \cite{Du} shows BER degradation because it requires more pilot symbols and thus more power. The scheme in \cite{Jeon} shows somewhat degraded PAPR reduction performance because some phase rotation vectors are made by linear combination of other phase rotation vectors, which generates highly correlated phase rotation vectors.

Also, several low-complexity SLM schemes which utilize the signals at an intermediate stage of IFFT have been proposed~\cite{Lim}, \cite{Ghassemi}. In those schemes, the signals at an intermediate stage of IFFT are multiplied by phase rotation vectors to generate alternative OFDM signal sequences, which can be equivalently viewed as multiplying phase rotation vectors to the input symbol sequence. Although these schemes give PAPR reduction performance close to that of the conventional SLM scheme without BER degradation, their computational complexity is still high.

In this paper, a low-complexity SLM scheme is proposed, which utilizes the signals at an intermediate stage of IFFT similar to \cite{Lim} and \cite{Ghassemi}. However, the proposed scheme generates each alternative OFDM signal sequence by cyclically shifting the connections in each subblock at an intermediate stage of IFFT. It can also be equivalently viewed as multiplying the corresponding phase rotation vectors which have lower correlations than those of \cite{Lim} and \cite{Ghassemi}, to the input symbol sequence. Consequently, the PAPR reduction performance of the proposed SLM scheme can approach to that of the conventional SLM scheme with lower computational complexity compared to the schemes in \cite{Lim} and \cite{Ghassemi}. Also, the proposed SLM scheme has no BER degradation compared to the conventional SLM scheme.

The rest of this paper is organized as follows. In Section II,
PAPR and the conventional SLM scheme are reviewed. In Section III,
a new low-complexity SLM scheme is proposed and analyzed. The proposed SLM scheme is evaluated through simulation in Section IV and conclusions are given in Section V.

\section{The Conventional SLM Scheme}

\begin{figure}[htbp]
\centering
\includegraphics[width=.9\linewidth]{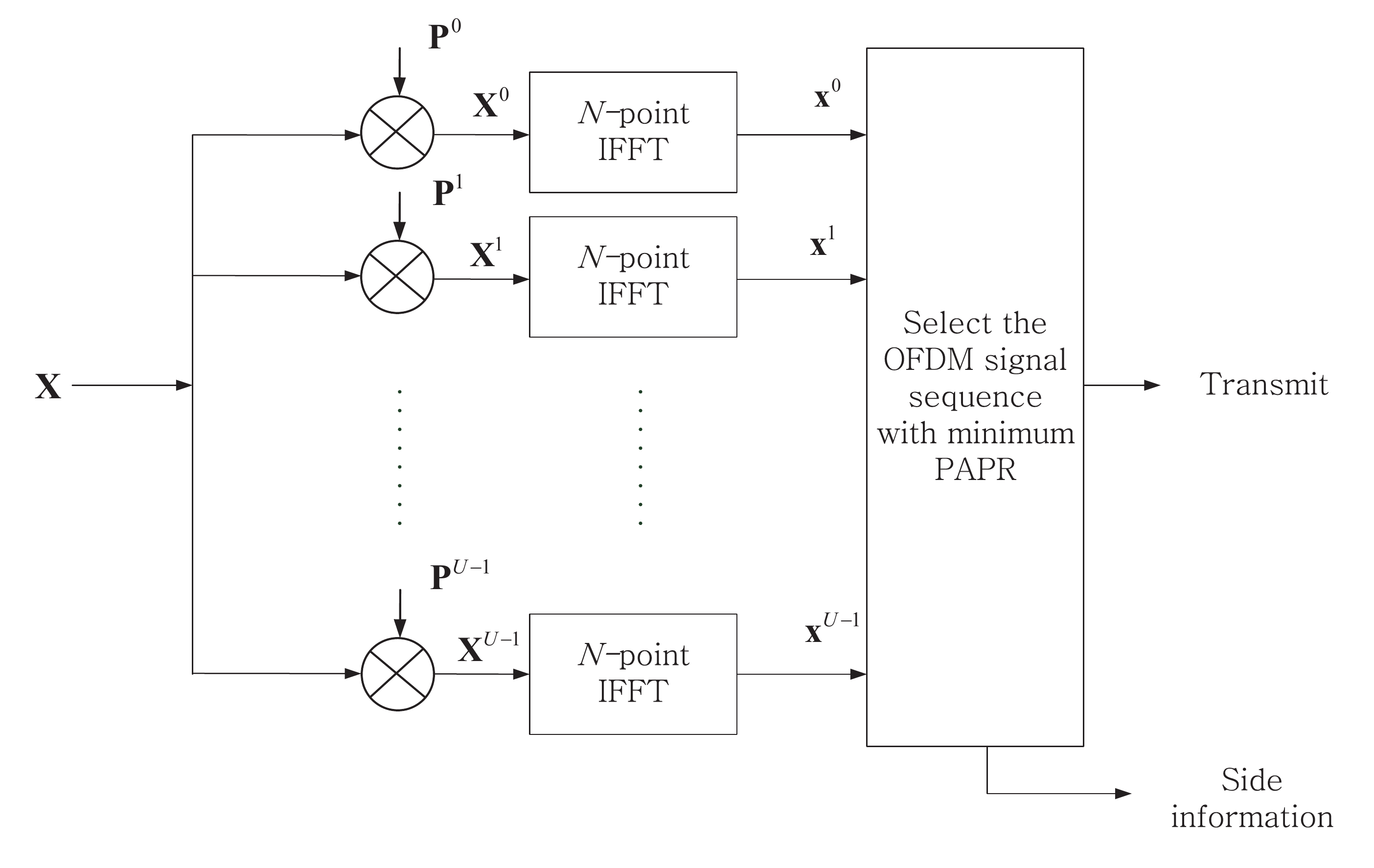}
\caption{A block diagram of the conventional SLM scheme.}
\label{fig:cslm}
\end{figure}

In this paper, we use the upper case $\mathbf{X}=\{X(0),X(1),...,X(N-1)\}$ for the input symbol sequence and the lower case $\mathbf{x}=\{x(0),x(1),...,x(N-1)\}$ for the OFDM signal sequence, where $N$ is the number of subcarriers. For simplicity, oversampling is not considered, but the proposed SLM scheme can be directly applied to the oversampled OFDM signals like other modified SLM schemes. The relation between the input symbol sequence $\mathbf{X}$ in frequency domain and the OFDM signal sequence $\mathbf{x}$ in time domain can be expressed by IFFT as
\begin{equation}\label{eq:IFFTeq}
x(n)=\sum_{k=0}^{N-1}X(k)W^{-kn}
\end{equation}
where $W=e^{-j\frac{2\pi}{N}}$ and $0 \leq n \leq N-1$.

The conventional SLM scheme in \cite{Bauml} is described in Fig.~\ref{fig:cslm}, which generates $U$
alternative OFDM signal sequences ${\mathbf x}^{u}$,
$0\le u \le U-1$, for the same input symbol sequence $\mathbf{X}$. To generate $U$ alternative OFDM signal sequences, $U$ distinct phase rotation vectors ${\mathbf P}^u$ known to both transmitter and receiver are used, where ${\mathbf P}^{u}$=$\{P^u(0),~P^u(1),\cdots,~P^u(N-1)\}$ with $P^u(k)=e^{j\phi^u(k)}$, $\phi^u(k)\in [0,~2\pi )$, $0\le u\le U-1$. Note that in general, each element of the phase rotation vector ${\mathbf P}^u$ is a power of the primitive $K$-th
root of unity $e^{j\frac{2\pi}{K}}$ and $K=2$ or $4$ is generally used. ${\mathbf P}^0$ is the all-one vector for generating the original OFDM signal sequence and thus $\mathbf{x}^0=\mathbf{x}$. An input symbol sequence ${\mathbf X}$ is multiplied by each phase rotation vector ${\mathbf P}^u$ element by element. Then an input symbol sequence ${\mathbf X}$
is represented by $U$ different alternative input symbol sequences ${\mathbf X}^u$, where ${X}^u(k)= { X}(k){ P}^u(k)$, $0\leq u\leq U-1$.
These $U$ alternative input symbol sequences are IFFTed to generate $U$ alternative OFDM signal seuences~${\mathbf x}^{u}= {\rm IFFT}( {\mathbf X}^u )$ and the PAPR values of them are calculated. Finally, the alternative OFDM signal sequence ${\mathbf x}^{\tilde u}$ having the
minimum PAPR is selected for transmission as
\begin{equation}
\tilde{ u}=\underset{\scriptscriptstyle 0\le u \le U-1}{\rm arg~min}
\Bigg(\frac{\mathrm{max}| {x}^u(n) | ^2}{E[| {x}^u(n) | ^2]}\Bigg).
\end{equation}
Note that the SI on $\tilde{ u}$ needs to be transmitted in order to properly demodulate the received OFDM signal sequence at the receiver.

\section{A New SLM Scheme With Low Complexity}

\subsection{A New SLM Scheme}

\begin{figure}[htbp]
\centering
\includegraphics[width=.6\linewidth]{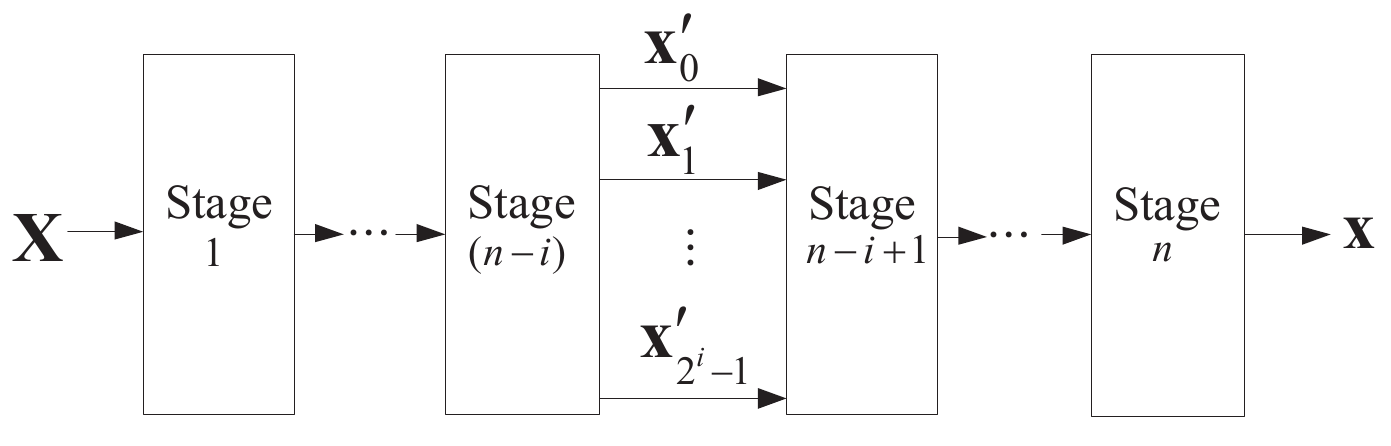}
\caption{A block diagram of the ordinary $N$-point decimation-in-frequency IFFT ($n=\mathrm{log}_2N$).}
\label{fig:IFFTblock}
\end{figure}
\begin{figure}[htbp]
\centering
\includegraphics[width=.9\linewidth]{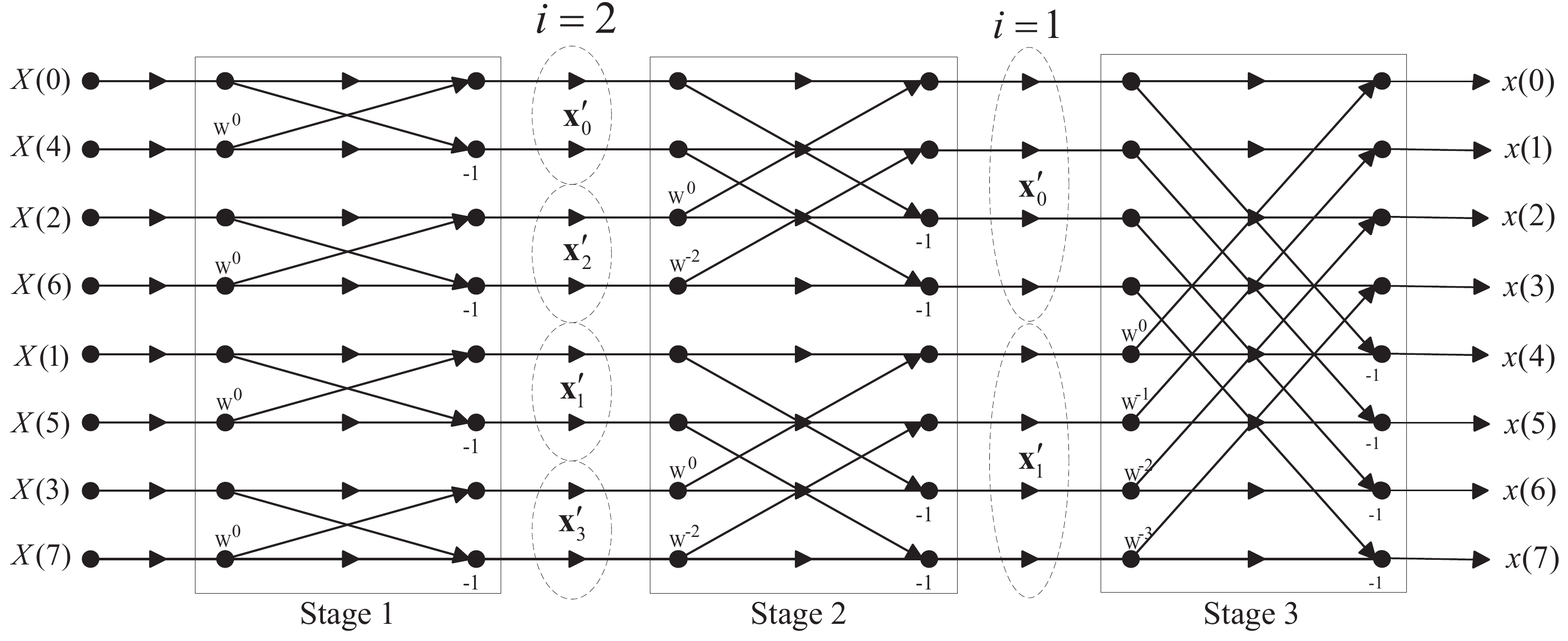}
\caption{Subblock partitions at stage 1 (i.e., $i=2$) and stage 2 (i.e., $i=1$) of IFFT when $N=8$ ($W=e^{-j\frac{2\pi}{8}}$).}
\label{fig:dififft}
\end{figure}

Prior to explaining the proposed SLM scheme, we describe the ordinary decimation-in-frequency IFFT structure. It is well known that the ordinary $N$-point decimation-in-frequency IFFT can be viewed as in Fig.~\ref{fig:IFFTblock}, where $n=\mathrm{log}_2N$. For any integer $i$, $1\leq i \leq n-1$, the intermediate OFDM signal sequence $\mathbf{x}'$ at stage $(n-i)$ is divided into $2^i$ subblocks $\mathbf{x}'_0,\mathbf{x}'_1,...,\mathbf{x}'_{2^i-1}$. A subblock $\mathbf{x}'_m$ is composed of $2^{n-i}$ outputs from the stage $(n-i)$ of IFFT, which is equivalent to the $2^{n-i}$-point IFFT using the input symbol sequence $X(k)$ satisfying $k~{\rm mod}~2^i=m$. Fig.~\ref{fig:dififft} shows an example of subblock partitions when $N=8$ and $i=1,~2$.

\begin{figure}[htbp]
\centering
\includegraphics[width=.999999\linewidth]{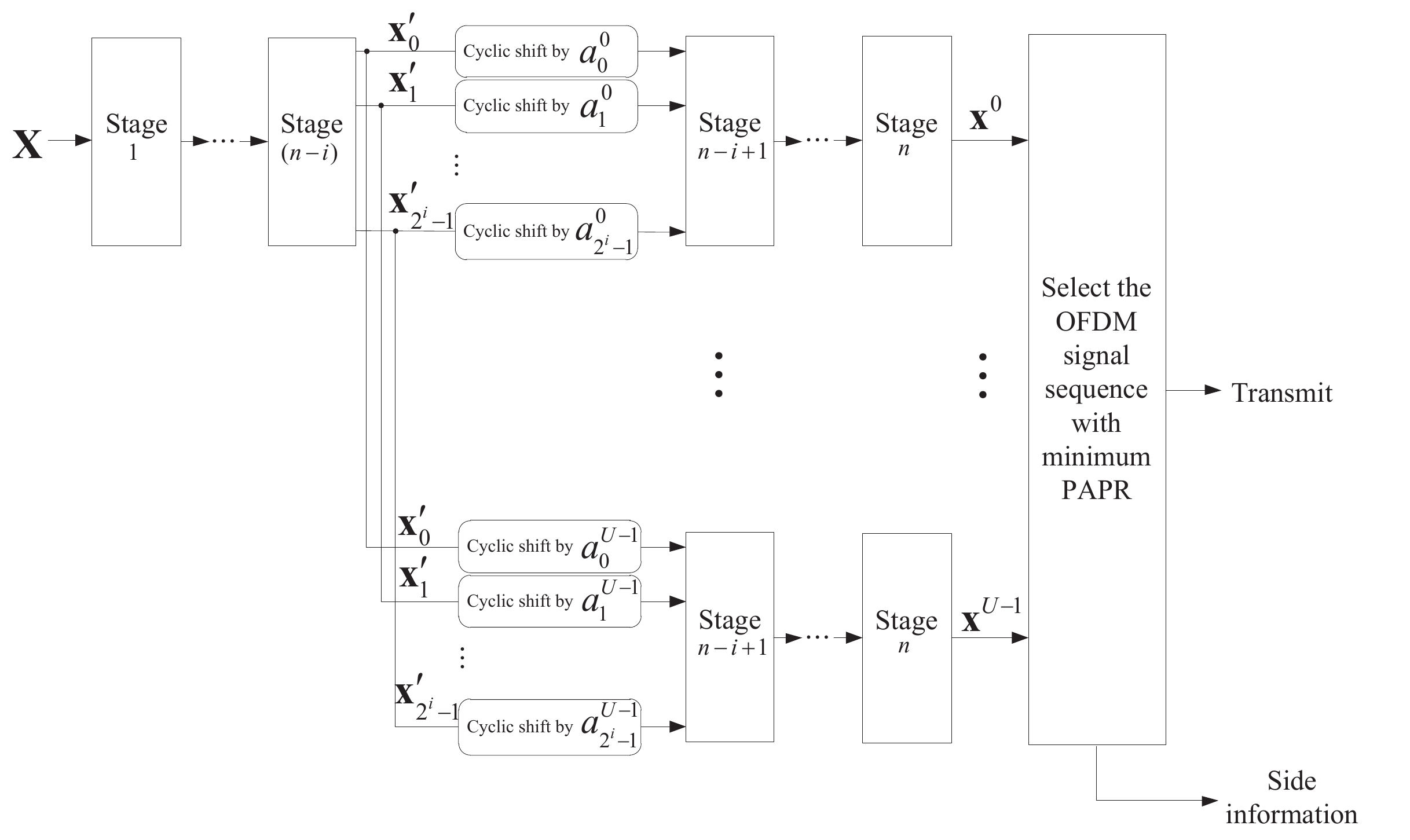}
\caption{A block diagram of the proposed SLM scheme ($n=\mathrm{log}_2N$).}
\label{fig:blockdiagram0}
\end{figure}
Fig.~\ref{fig:blockdiagram0} shows a block diagram of the proposed SLM scheme.
The $N$ input symbols $X(k)$, $0\leq k\leq N-1$, are processed by the ordinary $N$-point decimation-in-frequency IFFT up to the stage $(n-i)$, where $i$ is the number of remaining stages until finishing the IFFT.
To generate the $j$th alternative OFDM signal sequence, $0 \leq j \leq U-1$, the connections in each of subblocks $\mathbf{x}'_0,\mathbf{x}'_1,...,\mathbf{x}'_{2^i-1}$ are cyclically shifted upward by the predetermined integer numbers,
$a_0^j,a_1^j,...,a_{2^i-1}^j$, respectively.
Then these cyclically shifted $2^i$ subblocks become the input to the stage $(n-i+1)$ of $N$-point IFFT to generate the $j$th alternative OFDM signal sequence $\mathbf{x}^j$.
Finally, among these $U$ alternative OFDM signal sequences, the one having the minimum PAPR is selected for transmission and the SI is also transmitted.

\begin{figure}[htbp]
\centering
\includegraphics[width=.9\linewidth]{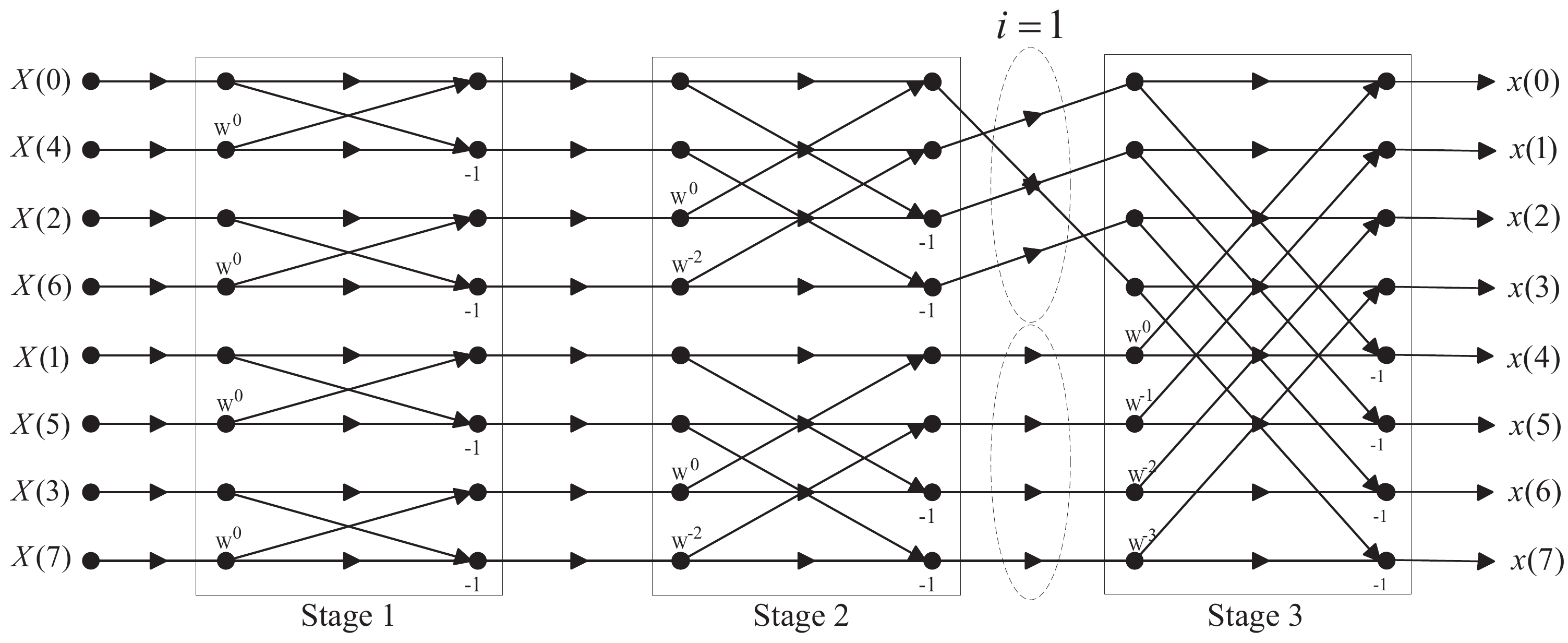}
\caption{An alternative OFDM signal sequence generated by the proposed scheme for $N=8$ and $i=1$ using $a_0=1$ and $a_1=0$.}
\label{fig:n=8}
\end{figure}
Fig.~\ref{fig:n=8} shows an example to generate an alternative OFDM signal sequence by the proposed scheme for $N=8$ and $i=1$ using $a_0=1$ and $a_1=0$. Clearly, the original OFDM signal sequence $\mathbf{x}^0$ is generated by using $a_0=0$ and $a_1=0$. Other alternative OFDM signal sequences are generated by simply changing the shift values $a_0$ and $a_1$.
For $i=2$, each of four subblocks, $\mathbf{x}'_0,\mathbf{x}'_1,\mathbf{x}'_2,\mathbf{x}'_3$ is cyclically shifted and the last two stages of $8$-point IFFT are performed as the ordinary IFFT.

The value $i$ can be any of $1,2,...,n-1$. As $i$ increases, the PAPR reduction performance improves but the computational complexity also increases, which will be explained in the following subsections.
Also, a selection method of shift values $a_0^j,a_1^j,...,a_{2^i-1}^j$ to achieve good PAPR reduction performance is analyzed and proposed in Subsections III-C and III-D. Compared with the conventional SLM scheme, the proposed scheme can substantially reduce the amount of computations for IFFTs to generate $U$ alternative OFDM signal sequences, which will be analyzed in Subsection III-E.

\subsection{Relation Between the Proposed SLM Scheme and the Conventional SLM Scheme}

In this subsection, the relation between the proposed SLM scheme and the conventional SLM scheme is investigated.
Let $M=2^i$ be the number of subblocks at the stage $(n-i)$ in the $N$-point decimation-in-frequency IFFT, where $N=2^n$ and $L=N/M$ is the size of each subblock.
Then, by replacing $k$ with $Ml+m$, (\ref{eq:IFFTeq}) can be rewritten as
\begin{align}\label{eq:proifft}
x(n)&=\sum_{m=0}^{M-1}\sum_{l=0}^{L-1}X(Ml+m)W^{-(Ml+m)n}\nonumber\\
&=\sum_{m=0}^{M-1}\Bigg(\sum_{l=0}^{L-1}X(Ml+m)W^{-Mln}\Bigg)W^{-mn}.
\end{align}
Note that $\sum_{l=0}^{L-1}X(Ml+m)W^{-Mln}, 0\leq m\leq M-1$, in (\ref{eq:proifft}) corresponds to the subblock $\mathbf{x}'_m$ of the intermediate OFDM signal sequence at the stage $(n-i)$.

The $j$th alternative OFDM signal sequence is generated by cyclically shifting the connections in each subblock $\mathbf{x}'_m$ by $a_m^j$ and processing the remaining stages of IFFT. Thus, the $j$th alternative OFDM signal sequence can be expressed as
\begin{align}\label{eq:proalter}
x^j(n)&=\sum_{m=0}^{M-1}\Bigg(\sum_{l=0}^{L-1}X(Ml+m)W^{-Ml(n+a^j_m)}\Bigg)W^{-mn}\nonumber\\
&=\sum_{m=0}^{M-1}\sum_{l=0}^{L-1}X(Ml+m)W^{-Mla^j_m}W^{-(Ml+m)n}.
\end{align}
By replacing $Ml+m$ with $k$ and noting that $m=k~\mathrm{mod}~M$ and $Ml=k-(k~\mathrm{mod}~M)$, the $j$th alternative OFDM signal sequence in (\ref{eq:proalter}) can be
expressed as
\begin{equation}
x^j(n)=\sum_{k=0}^{N-1}X(k)W^{-(k-(k~\textrm{mod}~M))a^j_{k\: \textrm{mod}\: M}}W^{-kn}.
\end{equation}

Clearly, the proposed SLM scheme can be equivalently viewed as the conventional SLM scheme using the phase rotation vectors given as
\begin{equation}\label{eq:eqprv}
P^j(k)=W^{-(k-(k~\textrm{mod}~M))a^j_{k\: \textrm{mod} \:M}}.
\end{equation}
Therefore, the receiver of the proposed SLM scheme is identical to that of the conventional SLM scheme. Since the $P^j(k)$ in (\ref{eq:eqprv}) has unit magnitude for all $j$ and $k$, the proposed SLM scheme does not degrade the BER performance as the conventional SLM scheme.

\subsection{Good Shift Values for the Proposed SLM Scheme}

It is clear that the shift values have a big impact on the PAPR reduction performance of the proposed scheme. It is well known that the optimal phase rotation vectors should be orthogonal and aperiodic for SLM scheme~\cite{Limphase}. However, for the correlated phase rotation vectors, the PAPR reduction performance can be analyzed by using the relation between the correlation of component powers of alternative OFDM signal sequences and the correlation of phase rotation vectors as in~\cite{Heo}.

Let $\mathrm{P_c}^j(n)$, $0 \leq n \leq N-1$, denote the $n$th component power $|x^j(n)|^2$ of the $j$th alternative OFDM signal sequence $\mathbf{x}^j$. In~\cite{Heo}, a design criterion of phase rotation vectors in SLM scheme with $U$ alternative OFDM signal sequences was derived by using the correlation coefficient $\rho_{jv}(\tau)$ between $\mathrm{P_c}^j(n)$ and $\mathrm{P_c}^v(n+\tau)$, $0\leq \tau \leq N-1$, where $0 \leq j \ne v \leq U-1$. It was also shown that the PAPR reduction performance improves as the maximum value of $\rho_{jv}(\tau)$ for $\tau$ decreases. As in~\cite{Heo}, $\rho_{jv}(\tau)$ can be approximated as
\begin{equation}\label{eq:corr}
\rho_{jv}(\tau)\simeq \frac{1}{N^2} \bigg| \sum_{k=0}^{N-1}P^j(k)P^v(k)^*W^{k\tau} \bigg| ^2
\end{equation}
where $(\cdot)^*$ denotes the complex conjugate. Therefore, to achieve good PAPR reduction performance, the shift values $\{a_0^j,a_1^j,...,a_{M-1}^j\}$ and $\{a_0^v,a_1^v,...,a_{M-1}^v\}$ should be chosen such that
\begin{equation}\label{eq:minmax}
\underset{\scriptscriptstyle a_0^j,a_1^j,...,a_{M-1}^j,a_0^v,a_1^v,...,a_{M-1}^v}{\rm arg~min}
\Bigg(\underset{\scriptscriptstyle \tau}{\rm max}
~\rho_{jv}(\tau)\Bigg)
\end{equation}
where $a_0^j,a_1^j,...,a_{M-1}^j,a_0^v,a_1^v,...,a_{M-1}^v\in\{0,1,...,L-1\}$.
For solving this problem, by replacing $k$ with $Ml+m$, we can rewrite (\ref{eq:corr}) as
\begin{equation}\label{eq:mlcorr}
\rho_{jv}(\tau)\simeq\frac{1}{N^2}\bigg| \sum_{m=0}^{M-1}\sum_{l=0}^{L-1}P^j(Ml+m)P^v(Ml+m)^*W^{(Ml+m)\tau}\bigg| ^2.
\end{equation}
By using $P^j(Ml+m)=W^{-Mla^j_m}$ in (\ref{eq:eqprv}), (\ref{eq:mlcorr}) can be given as
\begin{align}\label{eq:serieseasy}
\rho_{jv}(\tau)&\simeq\frac{1}{N^2}\bigg| \sum_{m=0}^{M-1}\sum_{l=0}^{L-1}W^{M(a^v_m-a^j_m+\tau)l+m\tau}\bigg| ^2\nonumber\\
&=\frac{1}{N^2}\bigg| \sum_{m=0}^{M-1}\frac{W^{m\tau}((W^{M(a_m^v-a_m^j+\tau)})^L-1)}{W^{M(a_m^v-a_m^j+\tau)}-1}\bigg| ^2\nonumber\\
&=\frac{1}{N^2}\bigg| A_0+A_1+...+A_{M-1} \bigg|^2
\end{align}
where
\begin{equation}
A_m=\frac{W^{m\tau}((W^{M(a_m^v-a_m^j+\tau)})^L-1)}{W^{M(a_m^v-a_m^j+\tau)}-1},~0\leq m\leq M-1.
\end{equation}

Since $ML=N$, the numerator of $A_m$ is always zero and thus $A_m$ is also zero except when the denominator of $A_m$ is zero.
When the denominator of $A_m$ is zero, it is easy to show that $A_m=LW^{m\tau}$.
The value of $\tau$ which generates nonzero $A_m$ can be found by solving
\begin{equation}
a^v_m-a^j_m+\tau=0~\mathrm{mod}~L.
\end{equation}
Since $-L < a^v_m-a^j_m < L$ and $0 \leq \tau < N$, the denominator of $A_m$ becomes zero if
\begin{equation}
\label{eq:tau}
\tau=\begin{cases} cL-(a^v_m-a_m^j),1\leq c \leq M, &a^v_m-a_m^j\geq 0\\
                   cL-(a^v_m-a_m^j),0\leq c \leq M-1, &a^v_m-a_m^j< 0.\end{cases}
\end{equation}
For each $m$, as the integer $\tau$ runs from $0$ to $N-1$, nonzero $A_m$ appears $M$ times.
Therefore, it is clear that $\underset{\scriptscriptstyle \tau}{\rm max}~\rho_{jv}(\tau)$ in (\ref{eq:minmax}) is minimized if $A_m$'s are not overlapped each other. In other words, it is required that at most one $A_m$ in (\ref{eq:serieseasy}) is nonzero for any $\tau$, which can be achieved if the following condition is satisfied;
\begin{center}
Condition for good shift values :\\
 For all $m_1\neq m_2$, $(a_{m_1}^v-a_{m_1}^j)-(a_{m_2}^v-a_{m_2}^j)\neq0~\mathrm{mod}~L$.
\end{center}

If this condition is satisfied, the maximum value of $\rho_{jv}(\tau)$ becomes $L^2/N^2$. If this condition is not satisfied for some $m$, the maximum value of $\rho_{jv}(\tau)$ becomes larger than $L^2/N^2$. For instance, suppose that $a_{m_1}^v-a_{m_1}^j=a_{m_2}^v-a_{m_2}^j=d>0$ for $m_1\neq m_2$ and the condition is satisfied for other $m$'s. Then, for $\tau=cL-d$, $1\leq c \leq M$, $\rho_{jv}(\tau)\simeq \frac{1}{N^2}|LW^{m_1\tau}+LW^{m_2\tau}|^2$ from (\ref{eq:serieseasy}) and it is easy to check that
\begin{equation}
\max_{1\leq c \leq M}\frac{1}{N^2}\mid LW^{m_1(cL-d)}+LW^{m_2(cL-d)} \mid^2 > \frac{L^2}{N^2}.
\end{equation}
Similarly, if there are more than two distinct $m$'s which do not satisfy the condition, it can be shown that the maximum value of $\rho_{jv}(\tau)$ is larger than $L^2/N^2$.

As a result, in order to achieve the best PAPR reduction performance of the proposed scheme with $U$ alternative OFDM signal sequences, shift values should satisfy the condition for good shift values for all $j,v$ pair, where $0 \leq j \ne v \leq U-1$. In this case, the maximum value of $\rho_{jv}(\tau)$ is $L^2/N^2$ for all $j,v$ pair. Hence, for the same $N$, the PAPR reduction performance of the proposed scheme improves as $i$ increases (i.e., $L^2/N^2$ decreases), which will be shown in Section IV.

\subsection{Methods to Generate Good Shift Values}

In this subsection, two methods to generate good shift values for the proposed SLM scheme are introduced. Firstly, random generation of shift values can be one of proper methods. If we choose $a_m^j$ for all $j$ and $m$ from $\{ 0,1,...,L-1 \}$ with equal probability $1/L$, then the term $(a_{m_1}^v-a_{m_1}^j)-(a_{m_2}^v-a_{m_2}^j)~\mathrm{mod}~L$ can take the value from $\{ 0,1,...,L-1 \}$ with equal probability. Therefore, shift values generated by random generation method satisfy the condition for good shift values with high probability because the practical value of $L$ is usually big. However, when we use the random generation method, both transmitter and receiver require the memory space to save $M(U-1)$ shift values (except $0$'s for the original OFDM signal sequence).

Secondly, we introduce a deterministic method to generate the shift values satisfying the condition for good shift values. We set $a_m^j = mj$, which is called $mj$-method. Then, $(a_{m_1}^v-a_{m_1}^j)-(a_{m_2}^v-a_{m_2}^j)$ can be rewritten as
\begin{align}\label{eq:mu}
(a_{m_1}^v-a_{m_1}^j)-(a_{m_2}^v-a_{m_2}^j)
&=(m_1v-m_1j)-(m_2v-m_2j)\nonumber\\
&=(m_1-m_2)(v-j).
\end{align}
Since we only consider the case when $0 \leq m_1 \ne m_2 \leq M-1$ and $0 \leq j\ne v \leq U-1$, we obtain
\begin{equation}\label{eq:muscope}
0<|(m_1-m_2)(v-j)| \leq (M-1)(U-1).
\end{equation}
From (\ref{eq:mu}) and (\ref{eq:muscope}), the $mj$-method is guaranteed to satisfy the condition for good shift values when $(M-1)(U-1)<L$, i.e. $(2^i-1)(U-1)< 2^{n-i}$. This inequality can be satisfied for practical value of $n$ and $U$ because the appropriate value of $i$ is $2$ or $3$ in the proposed scheme as will be shown in later section. Besides, the $mj$-method does not require the memory space to save the shift values, which is an additional advantage of the proposed SLM scheme using the $mj$-method compared to other SLM schemes requiring memory space to save the phase rotation vectors.

\subsection{Computational Complexity of the Proposed SLM Scheme}

In this subsection, the computational complexity of the proposed scheme is compared with those of the conventional SLM scheme and other low-complexity SLM schemes. We only compare the computational complexity to generate
alternative OFDM signal sequences because the remaining
computational complexity is the same for most SLM schemes if the number of alternative OFDM signal sequences is the same.

When the number of subcarriers is $N=2^n$, the numbers of complex multiplications and
complex additions required for the conventional SLM scheme can be derived as follows. An $N$-point IFFT
requires $(N/2)\textrm{log}_2N$ complex multiplications and $N\textrm{log}_2N$
complex additions. Therefore, the total numbers of complex multiplications and
complex additions for the conventional SLM scheme using $U$ alternative OFDM signal sequences are $U(N/2)\textrm{log}_2N$ and $UN\textrm{log}_2N$, respectively.
In the proposed scheme, if the cyclic shifts are performed at the stage $(n-i)$, the numbers of required complex multiplications and complex additions are $(N/2)\textrm{log}_2N+(U-1)(i/n)(N/2)\textrm{log}_2N$ and $N\textrm{log}_2N+(U-1)(i/n)N\textrm{log}_2N$, respectively. Note that the reduction ratio of complex multiplications is the same as that of complex additions.
Therefore, the computational complexity reduction ratio (CCRR) of the proposed scheme over
the conventional SLM scheme is derived only for complex multiplication as
\begin{align}
\textrm{CCRR}&=\bigg(1-\frac{\textrm{Complexity~of~the~proposed~scheme}}{\textrm{Complexity~of~the~conventional~SLM}}\bigg)\times100~(\%) \nonumber\\
&=\bigg( 1-\frac{n+(U-1)i}{nU} \bigg) \times100~(\%)=\frac{(n-i)(U-1)}{nU}\times100~(\%).
\end{align}

\begin{table}
\caption{CCRR($\%$) of the proposed scheme compared to the conventional SLM.}
\vspace{2pt}\label{table:ccrr} \centering
\begin{tabular}{|c||c|c|c|c|c|c|c|c|c|}
\hline
         $N$ &           \multicolumn{ 3}{c|}{64} &           \multicolumn{ 3}{c|}{256} &          \multicolumn{ 3}{c|}{1024} \\
\hline
         $U$ &          4 &          8 &         16 &          4 &          8 &         16 &          4 &          8 &         16 \\
\hline
\hline
       $i=1$ &      62.5  &      72.9  &      78.1  &      65.6  &      76.6  &      82.0  &      67.5  &      78.8  &      84.4  \\
\hline
       $i=2$ &      50.0  &      58.3  &      62.5  &      56.3  &      65.6  &      70.3  &      60.0  &      70.0  &      75.0  \\
\hline
       $i=3$ &      37.5  &      43.8  &      46.9  &      46.9  &      54.7  &      58.6  &      52.5  &      61.3  &      65.6  \\
\hline
       $i=4$ &      25.0  &      29.2  &      31.3  &      37.5  &      43.8  &      46.9  &      45.0  &      52.5  &      56.3  \\
\hline
\end{tabular}

\end{table}

As shown in Table~\ref{table:ccrr}, the proposed scheme has much lower computational complexity than the conventional SLM scheme. For example, when $i=3, N=1024$,~and $U=8$, the computational complexity of the proposed scheme reduces by $61.3$$\%$ compared with the conventional SLM scheme with almost the same PAPR reduction performance. It is clear that the CCRR is large when $N$ is large and $i$ is small. However, for small $i$, there appears a large amount of degradation in the PAPR reduction performance compared to the conventional SLM scheme as will be shown in Section IV.
\begin{figure}[H]
\centering
\includegraphics[width=.9\linewidth]{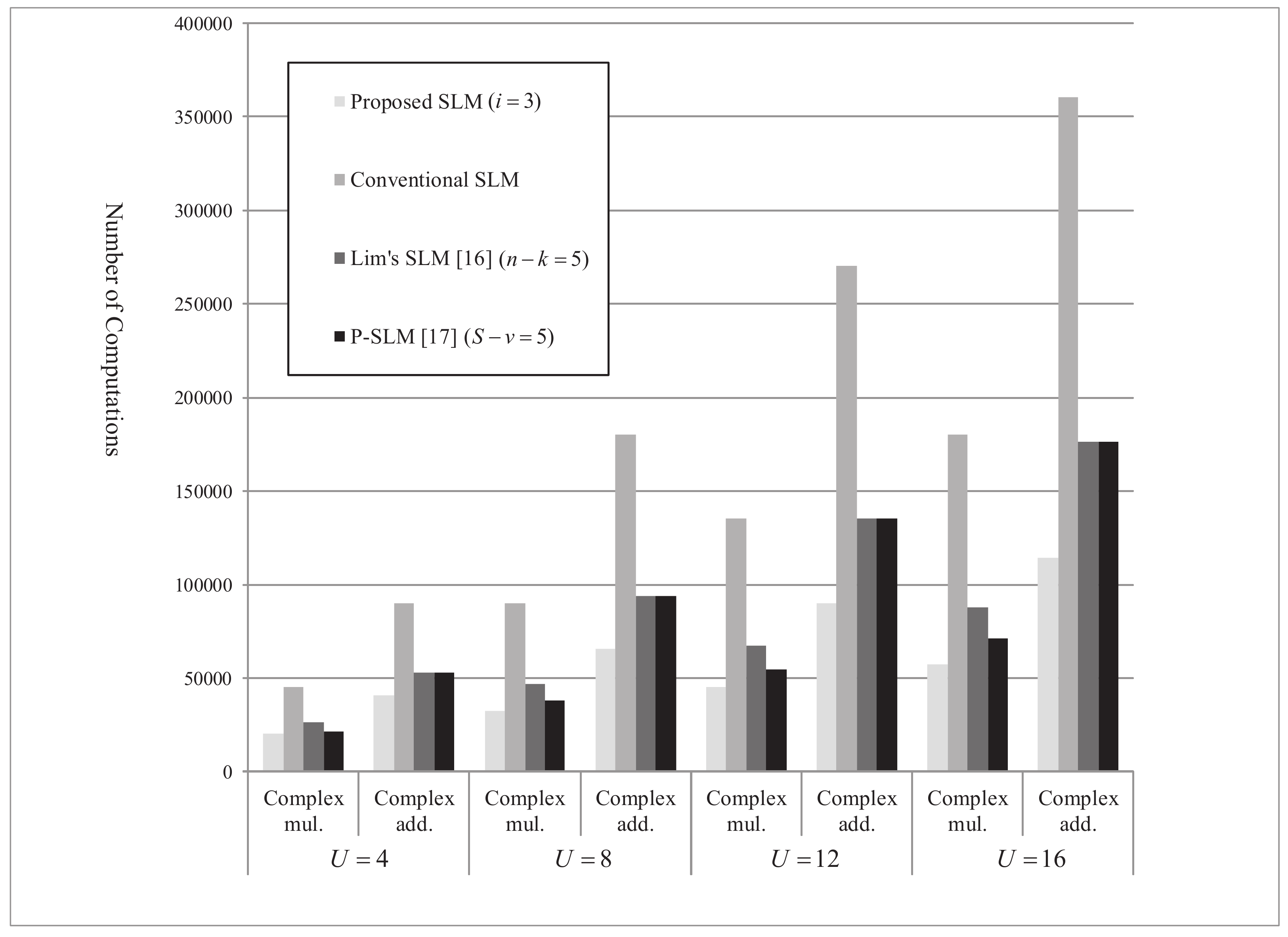}
\caption{Comparison of the computational complexity of the proposed SLM, P-SLM~\cite{Ghassemi}, Lim's SLM~\cite{Lim}, and the conventional SLM when $N = 2048$.}
\label{fig:lcSLM}
\end{figure}
Now, we compare the computational complexity of the existing low-complexity SLM schemes exploiting the signals at an intermediate stage of IFFT. The reason for this comparison is that their PAPR reduction performance is generally almost the same as that of the conventional SLM scheme with the same number of alternative OFDM signal sequences, which is different from most of other low-complexity SLM schemes.
Fig.~\ref{fig:lcSLM} shows the comparison of the computational complexity of the proposed SLM scheme, conventional SLM scheme, Lim's SLM scheme~\cite{Lim}, and P-SLM scheme~\cite{Ghassemi}. We set each low-complexity scheme to have the PAPR reduction performance close to that of the conventional SLM
scheme when $N = 2048$ and 16-quadrature amplitude modulation (16-QAM) is used. Fig.~\ref{fig:lcSLM} shows that the proposed SLM scheme has the lowest computational complexity among these SLM schemes.

\section{Simulation Results}

For the simulation, $10^6$ input symbol sequences are randomly generated and 16-QAM is used. For the conventional SLM scheme, each element of the phase rotation vectors is randomly selected from $\{\pm 1, \pm j\}$. Similarly, to determine the shift values for the proposed SLM scheme, random generation method is used. Note that random generation method and the $mj$-method show almost the same PAPR reduction performance for the practical values of $N$, $U$, and $i$ as will be shown in this section. However, in practical systems, the $mj$-method would be preferred because it does not require memory space to save the shift values.

\begin{figure}
\centering
\subfigure[]{
\includegraphics[width=.9\linewidth]{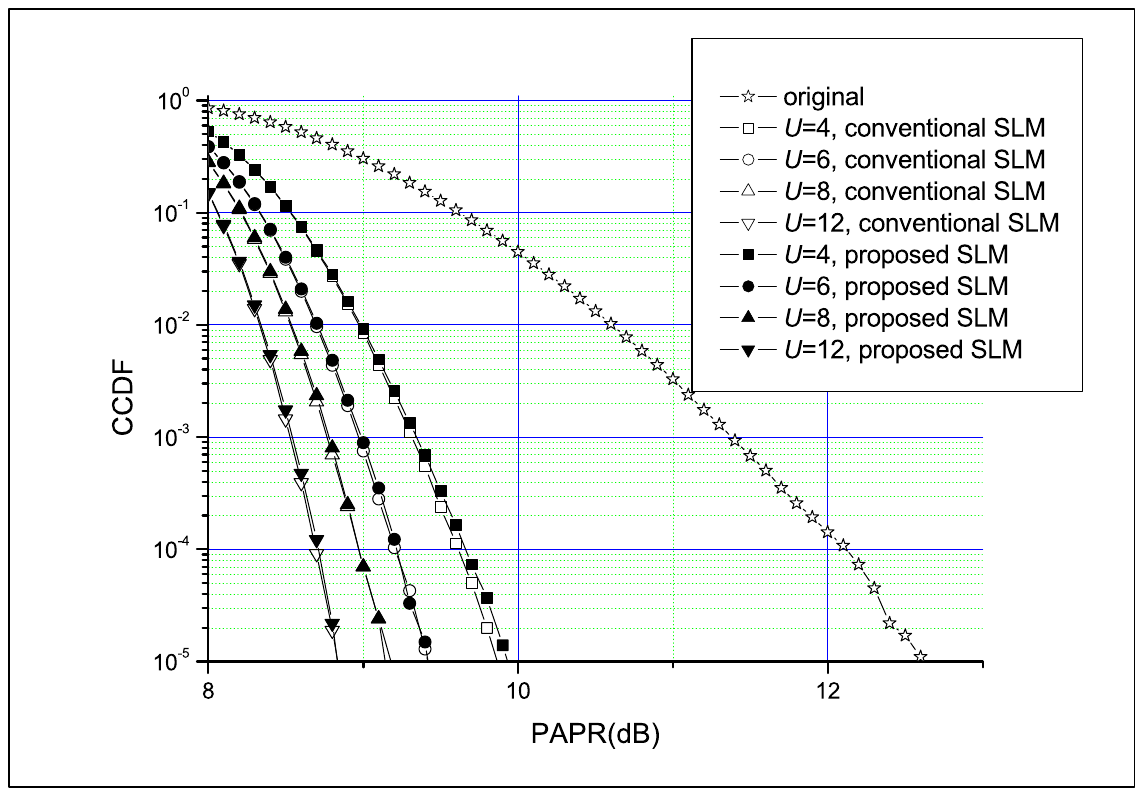}
}
\caption{Comparison of PAPR~reduction~performance~of~the~proposed and the conventional~SLM~schemes~when~$N=1024$~and~16-QAM~is~used~(a)~$i=1$,~(b)~$i=2$,~(c)~$i=3$.}
\label{fig:n1024i123}
\end{figure}

Fig.~\ref{fig:n1024i123} compares the PAPR reduction performance of the proposed SLM scheme with that of the conventional SLM scheme when $N=1024$ and 16-QAM is used for $i=1,2,3$. Fig.~\ref{fig:n1024i123} shows that the PAPR reduction performance of the proposed SLM scheme becomes better as $i$ increases, as expected from the analytical result that the maximum correlation coefficient value for the equivalent phase rotation vectors decreases as $i$ increases. It is also observed from Fig.~\ref{fig:n1024i123} that the PAPR reduction performance of the proposed SLM scheme becomes closer to that of the conventional SLM scheme as $i$ increases. When $i=3$, both schemes show almost the same PAPR reduction performance. Since the performance of the proposed SLM scheme is lower bounded by that of the conventional SLM scheme and the computational complexity increases as $i$ increases, the appropriate value of $i$ can be $2$ or $3$.

\begin{figure}
\centering
\subfigure[]{
\includegraphics[width=.9\linewidth]{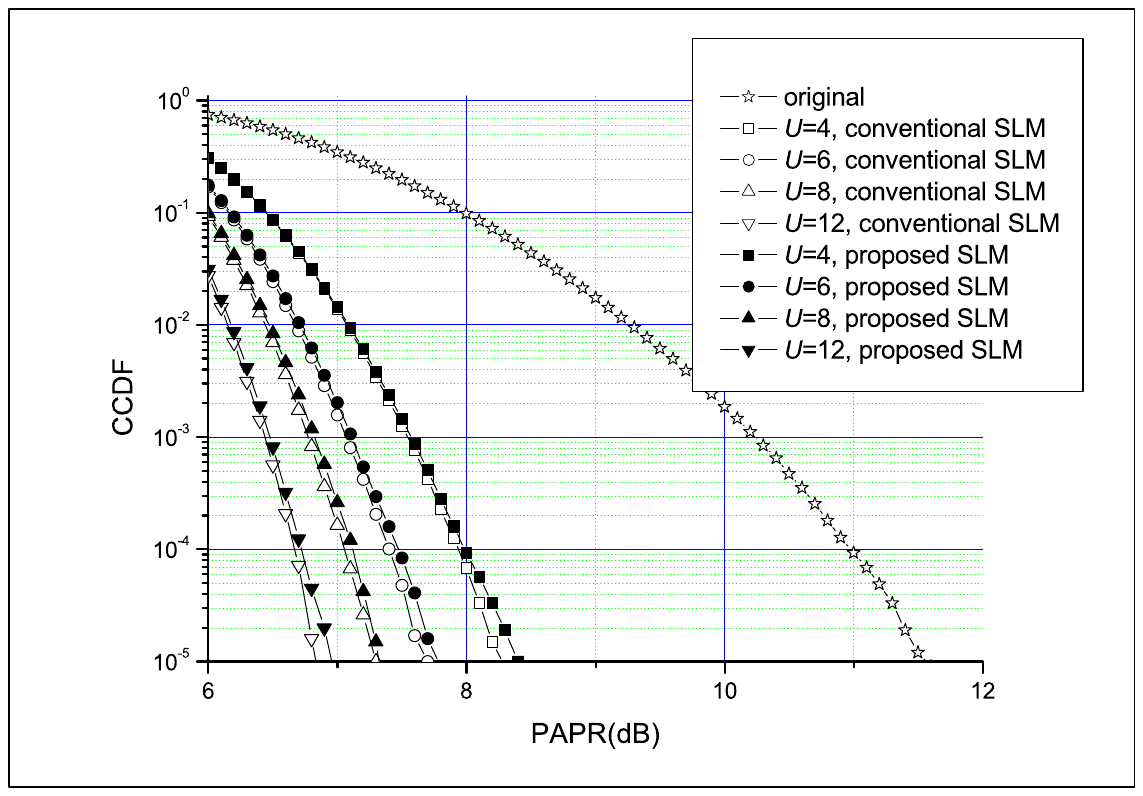}
}
\caption{Comparison of PAPR~reduction~performance~of~the~proposed and the conventional~SLM~schemes~when~$N=64$~and~16-QAM~is~used~(a)~$i=1$,~(b)~$i=2$,~(c)~$i=3$.}
\label{fig:n64i123}
\end{figure}

Fig.~\ref{fig:n64i123} compares the PAPR reduction performance of the proposed SLM scheme with that of the conventional SLM scheme when $N=64$ and 16-QAM is used for $i=1,2,3$. Fig.~\ref{fig:n64i123} also shows the similar tendency as Fig.~\ref{fig:n1024i123} and as $N$ increases, both schemes show closer performance.

\begin{figure}[!h]
\centering
\includegraphics[width=.9\linewidth]{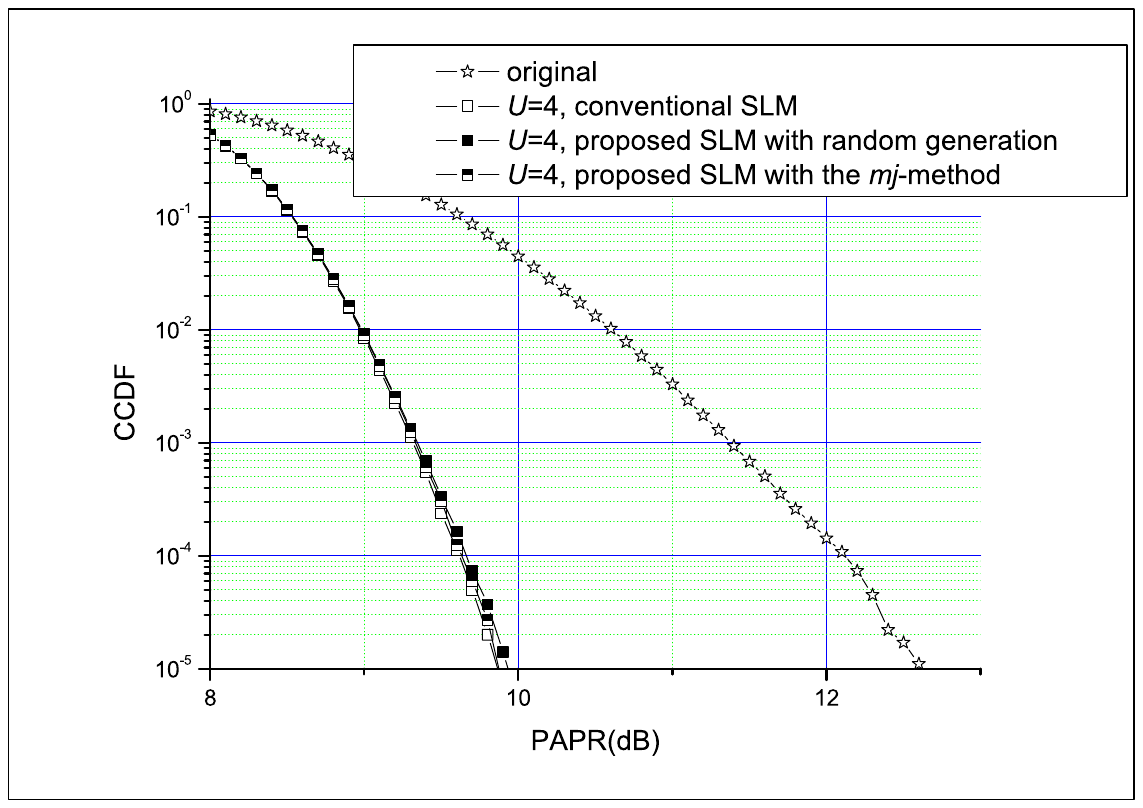}
\caption{Comparison~of~the~PAPR~reduction~performance~of~the~proposed~SLM~scheme~using the~$mj$-method~and~random~generation~method
~when~$N=1024$,~$U=4$,~$i=3$,~and~16-QAM~is~used.}
\label{fig:muvsran}
\end{figure}

Fig.~\ref{fig:muvsran} compares the PAPR reduction performance of the proposed SLM scheme using the $mj$-method and random generation method for selecting shift values. Since they show almost the same PAPR reduction performance, we can expect that two methods show almost the same PAPR reduction performance for practical values of $N$, $U$, and $i$. However, the $mj$-method requires no memory space to save the shift values (i.e., $U-1$ phase rotation vectors), which is different from other SLM schemes.

\section{Conclusions}

A new low-complexity SLM scheme exploiting the signals at an intermediate stage of IFFT is proposed, which shows almost the same PAPR reduction performance as the conventional SLM scheme when $i=3$. Instead of performing $U$ IFFTs as in the conventional SLM scheme, the proposed scheme operates one IFFT up to $(n-i)$ stages, which is common to generating all alternative OFDM signal sequences. Then, the connections in each subblock at the stage $(n-i)$ of IFFT is cyclically shifted by the predetermined shift value in the proposed SLM scheme. Since the cyclic shifts at an intermediate stage of IFFT can be viewed as multiplying an equivalent phase rotation vector which has unit magnitudes to the input symbol sequence, there is no BER degradation compared to the conventional SLM scheme.

The simulation results show that the proposed SLM scheme using $i=3$ can achieve almost the same PAPR reduction performance as the conventional SLM scheme. Also, it is verified that the proposed SLM scheme has the lowest computational complexity among existing low-complexity SLM schemes exploiting the signals at an intermediate stage of IFFT.

\section*{Acknowledgment}
This work was supported by the National Research Foundation of Korea (NRF) grant funded by the Korea government (MEST) (No. 2011-0000328).

\begin{figure}[]
\centering
\subfigure[]{
\includegraphics[width=.9\linewidth]{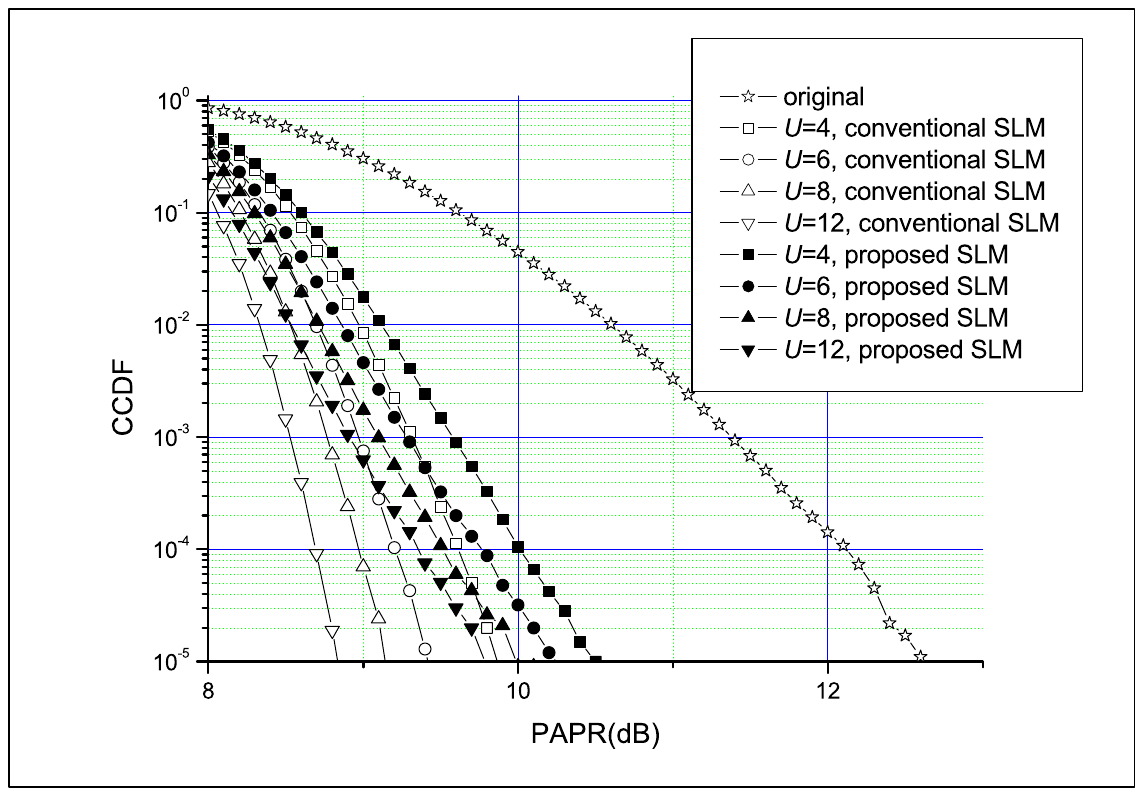}
}
\end{figure}

\begin{figure}[]
\subfigure[]{
\includegraphics[width=.9\linewidth]{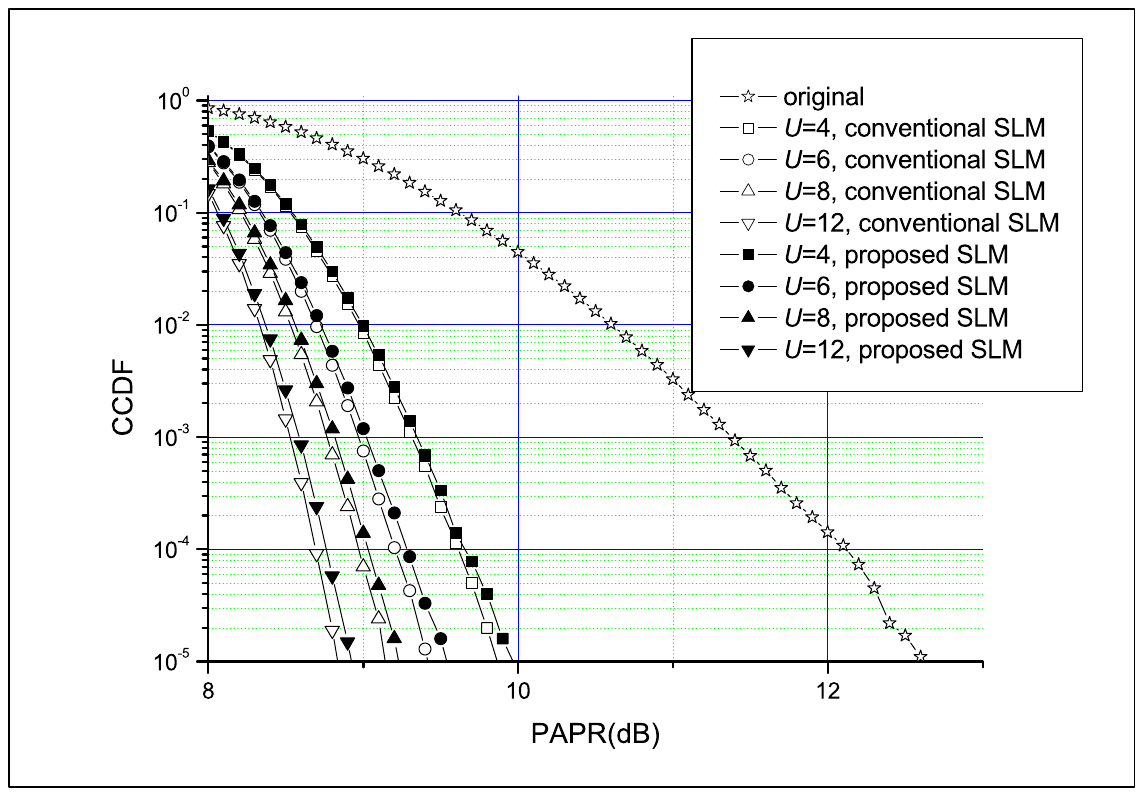}
}
\end{figure}

\begin{figure}[]
\centering
\subfigure[]{
\includegraphics[width=.9\linewidth]{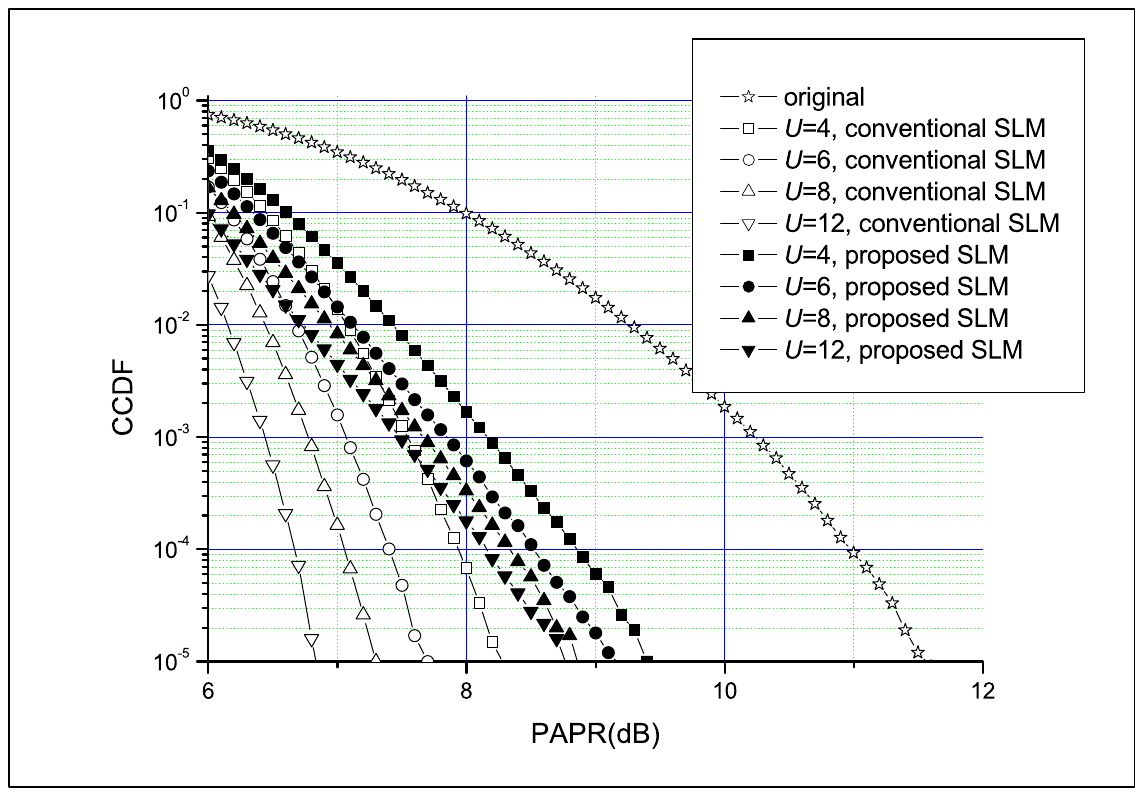}
}
\end{figure}

\begin{figure}[]
\subfigure[]{
\includegraphics[width=.9\linewidth]{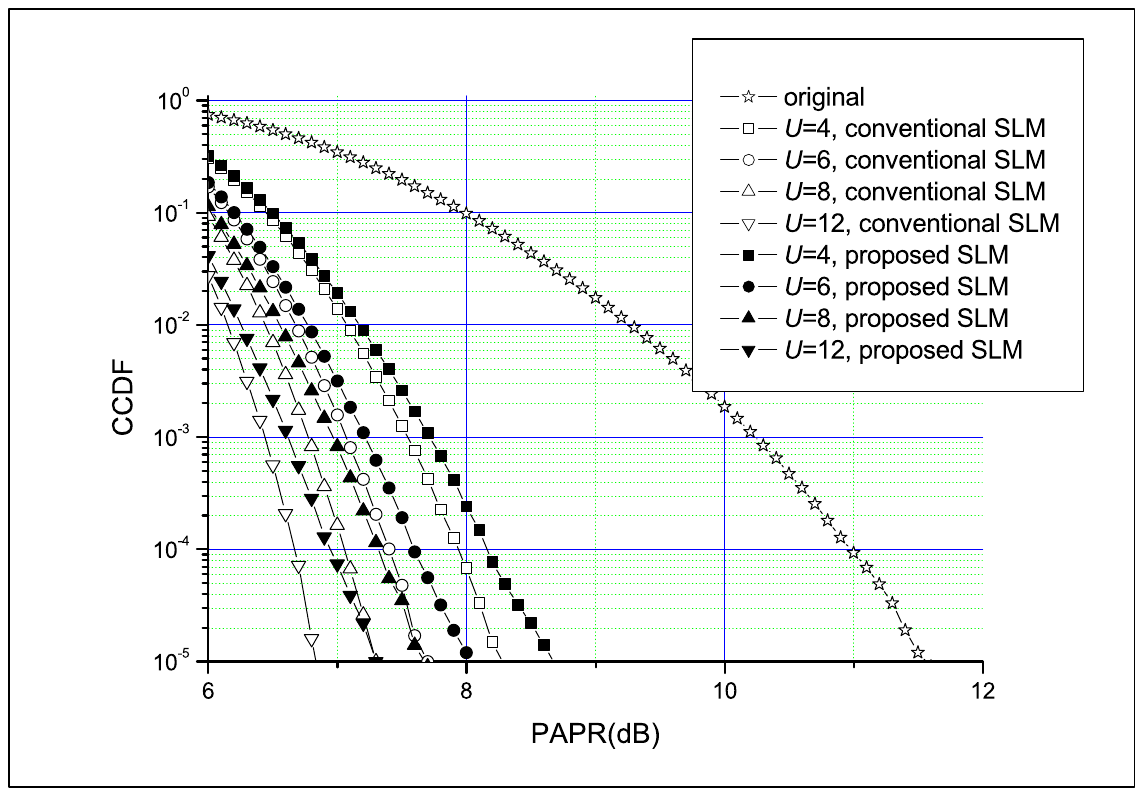}
}
\end{figure}

\begin{thebibliography}{99}

\bibitem{Oneal} R. O¡¯neal and L. N. Lopes, ``Envelope variation and spectral splatter
in clipped multicarrier signals,'' in \textit{Proc. IEEE PIMRC}, Sep. 1995, pp. 71--75.

\bibitem{YWang} L. Wang and C. Tellambura, ``Analysis of clipping noise and tone-reservation algorithms for peak reduction in OFDM systems,'' \textit{IEEE Trans. Veh.
Technol.}, vol. 57, no.3, pp. 1675--1694, May 2008.

\bibitem{Tsai} Y.-C. Tsai, S.-K. Deng, K.-C. Chen, and M.-C. Lin, ``Turbo coded OFDM for reducing PAPR and error rates,'' {\em IEEE Trans. Wireless Commun.}, vol. 7, no. 1, pp. 84--89, Jan. 2008.

\bibitem{Krongold} B. S. Krongold and D. L. Jones, ``PAR reduction in OFDM via active
constellation extension,'' {\em IEEE Trans. Broadcast.}, vol. 49, no. 3, pp.
258--268, Sep. 2002.

\bibitem{Tellado} J. Tellado and J. M. Cioffi, \textit{Multicarrier Modulation With Low PAR,
Application to DSL and Wireless.} Norwell, MA: Kluwer Academic Publisher, 2000.

\bibitem{muller} S.~H. M${\rm\ddot{u}}$ller, R. W. B${\rm\ddot{a}}$uml, R. F. H. Fischer, and J. B. Huber, ``OFDM with reduced peak-to-average power ratio by multiple signal
representation,'' {\em Ann. Telecommun.}, vol. 52,
no.1--2, pp. 58--67, Feb. 1997.

\bibitem{Yang} M. R. D. Rodrigues and I. J. Wassell, ``IMD reduction with SLM and PTS to improve the
error-probability performance of nonlinearly distorted OFDM signals,'' \textit{IEEE Trans. Veh.
Technol.}, vol. 55, no.2, pp. 537--548, Mar. 2006.

\bibitem{Bauml}
R. W. B${\rm\ddot{a}}$uml, R. F. H. Fischer, and J. B. Huber, ``Reducing the peak-to-average
power ratio of multicarrier modulation by selected mapping,'' \textit{Electron. Lett.}, vol. 32, no. 22, pp. 2056--2057, Oct. 1996.

\bibitem{Goff} S. Y. Le Goff, B. K. Khoo, C. C. Tsimenidis, and B. S. Sharif, ``A novel
selected mapping technique for PAPR reduction in OFDM systems,''
\textit{IEEE Trans. Commun.}, vol. 56, no. 11, pp. 1775--1779, Nov. 2008.

\bibitem{Limphase} D.-W. Lim, S.-J. Heo, J.-S. No, and H. Chung, ``On the phase sequences of SLM OFDM system for PAPR reduction,'' in {\em Proc. ISITA 2004}, Parma, Italy, Oct. 2004, pp. 230--235.

\bibitem{Heo} S.-J. Heo, H.-S. Joo, J.-S. No, D.-W. Lim, and D.-J. Shin, ``Analysis of PAPR reduction performance of SLM schemes with correlated phase vectors,'' in \textit{Proc. IEEE ISIT}, Seoul, Korea, Jun. 2009, pp. 1540--1543.

\bibitem{Wang0} C.-L. Wang and Y. Ouyang, ``Low-complexity selected mapping schemes for peak-to-average power ratio reduction in OFDM systems,'' \textit{IEEE Trans. Signal Process.}, vol. 53, no. 12, pp. 4652--4660, Dec. 2005.

\bibitem{Wang} C.-L. Wang and S.-J. Ku, ``Novel conversion matrices for simplifying
the IFFT computation of an SLM-based PAPR reduction scheme for
OFDM systems,'' \textit{IEEE Trans. Commun.}, vol. 57, no. 7, pp. 1903--1907, Jul. 2009.

\bibitem{Du} Z. Du, N. C. Beaulieu, and J. Zhu, ``Selected time-domain filtering
for reduced-complexity PAPR reduction in OFDM,'' \textit{IEEE Trans. Veh.
Technol.}, vol. 60, no.3, pp. 1170--1176, Mar. 2009.

\bibitem{Jeon} H.-B. Jeon, J.-S. No, and D.-J. Shin, ``A low-complexity
SLM scheme using additive mapping sequences for PAPR reduction of OFDM signals,'' accepted for publication in \textit{IEEE Trans. Broadcast.}, Apr. 2011.

\bibitem{Lim} D.-W. Lim, J.-S. No, C.-W. Lim, and H. Chung, ``A new SLM OFDM scheme with low complexity for PAPR reduction,''
{\em IEEE Signal Process. Lett.}, vol. 12, no. 2, pp. 93--96, Feb. 2005.

\bibitem{Ghassemi} A. Ghassemi and T. A. Gulliver, ``Partial selective mapping OFDM
with low complexity IFFTs,'' \textit{IEEE Comm. Lett.}, vol. 12, no. 1, pp. 4--6, Jan. 2008.

\bibitem{Kim} K.-H. Kim, H.-B. Jeon, J.-S. No, and D.-J. Shin , ``A new selected mapping scheme for PAPR reduction in OFDM systems,'' in \textit{Proc. IEEE ISITA 2010}, Taichung, Taiwan, Oct. 2010, pp.1054--1057.

\end{thebibliography}
\end{document}